\RequirePackage{ifpdf}
\RequirePackage{color}
\documentclass{JINST}

\title{Enhanced Photon Traps for Hyper-Kamiokande}

\author{Carsten Rott, Seongjin In \\ 
Department of Physics, Sungkyunkwan University, Suwon 440-746, Korea\\
 E-mail: \email{rott@skku.edu}}
\author{Fabrice Reti\`{e}re, Peter Gumplinger \\
  TRIUMF\\
  4004 Wesbrook Mall Vancouver BC V6T 2A3, Canada\\
  E-mail: \email{fretiere@triumf.ca}}

\abstract{Hyper-Kamiokande, the next generation large water Cherenkov detector in Japan, is planning to use approximately 80,000 20-inch photomultiplier tubes (PMTs). They are one of the major cost factors of the experiment. We propose a novel enhanced photon trap design based on a smaller and more economical PMT in combination with wavelength shifters, dichroic mirrors, and broadband mirrors. GEANT4 is utilized to obtain photon collection efficiencies and timing resolution of the photon traps. We compare the performance of different trap configurations and sizes. Our simulations indicate an enhanced photon trap with a 12-inch PMT can match a 20-inch PMT's collection efficiency, however at a cost of reduced timing resolution. The photon trap might be suitable as detection module for the outer detector with large photo coverage area.
}


\begin{document}

\section{Introduction}

The remarkable success of the 50-kton~(kilotons) Super-Kamiokande (Super-K)~\cite{Fukuda:2002uc} detector has sparked interests in novel large volume neutrino detectors. Hyper-Kamiokande (Hyper-K) represents the next generation of the highly-successful water Cherenkov technology employed for Super-K and foresees the construction of two large detectors. Two caverns will be excavated to each host a detector with a physical volume of 0.258~Mton~(Megatons), fiducial volume of 0.187~Mton. Detectors will have a 40\% photo coverage achieved by  40,000~inward looking 20-inch photomultiplier tubes (PMTs)~\cite{HK_Design,Abe:2015zbg,Abe:2011ts}. The primary candidate site is the Toshibura mine in Japan. An option of placing one of the detectors in Korea at the second oscillation maximum of the high intensity JPARC~(Japan Proton Accelerator Research Complex) beam is expected to enhance CP violation sensitivity and is actively being studied~\cite{T2HKK}. A major cost factor of Hyper-K and other next-generation large volume detectors are their photosensors. Smaller PMTs combined with enhanced photo sensing techniques through the usage of wavelength shifters (WLS) and dichroic mirrors might offer more economical alternatives with similar performance. 
 
Like Super-K, Hyper-K's baseline design relies on 20-inch PMTs at its inner detector to detect Cherenkov radiation. The PMTs are distributed evenly on a one-by-one meter area under water. Despite this high photo coverage, 83\% of the Cherenkov radiation reaching the inner detector PMT level is lost. A large fraction of the missed Cherenkov radiation however could be captured with a photon trap and guided to the PMT.
 
In our study we compare a 20-inch PMT with designs using 12-inch PMTs for reduced hardware costs by taking advantage of a wavelength shifter as a light trap. We characterize the performance of the different designs with respect to their photon collection and detection efficiency as well as their time-response. We show that the new designs proposed here offer interesting alternatives and should be seriously considered for next-generation detectors. 

Hyper-K is a multipurpose experiment that is designed to address a broad range of fundamental physics questions, with each of them posing its own detector requirements. We carry out a general performance evaluation of various enhanced photon trap designs and leave more detailed studies to experimental collaborations. Hyper-K's science objectives include the measurement of diffuse supernova neutrino background~\cite{Beacom:2010kk,Beacom:2003nk}, core-collapse supernova burst neutrinos~\cite{Hirata:1987hu,Ikeda:2007sa}, and indirect searches for dark matter~\cite{Rott:2012qb,Bernal:2012qh,Rott:2015nma,Murase:2016nwx}. Other physics objectives include proton decay~\cite{Shiozawa:1998si}, neutrino oscillation measurements~\cite{Fukuda:1998mi}, and CP violation with an upgraded J-PARC beam~\cite{Abe:2015zbg}. Further applications are in Earth science~\cite{Rott:2015kwa} and solar neutrinos~\cite{Hosaka:2005um,Cravens:2008aa,Abe:2010hy}. 

\begin{figure}[h]
	\centering
	\includegraphics[width=0.9\textwidth]{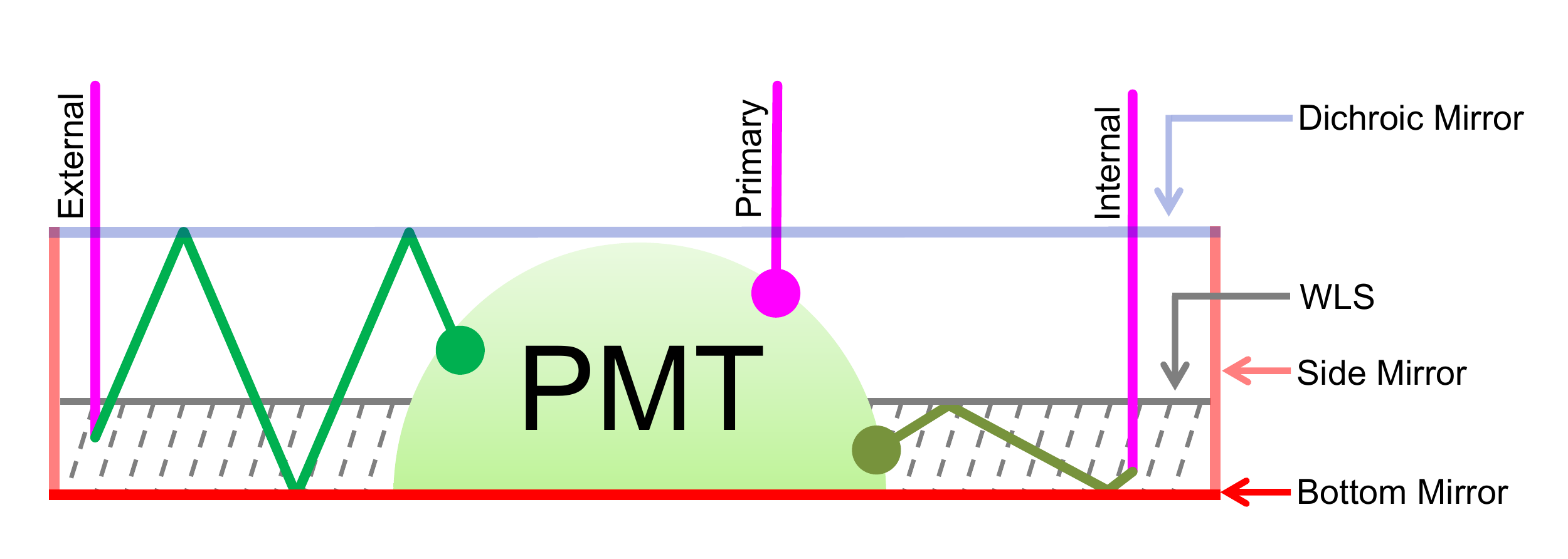}
	\caption{Classification of photon collection channels based on the light path taken in the photon trap. The example setup shown is the full trap (case~4), see text for description. The full trap is composed of broadband mirrors (red), dichroic mirror (blue) and wavelength shifter (grey). Events are classified with three different collection channels: {\it Primary}, {\it Internal} and {\it External} collection.}
	\label{fig_setup}
\end{figure}

Wavelength shifter plates have a long history of applications in high-energy physics (see for example~\cite{Botner:1980uf,Balka:1987ty}) and are also commonly used as photon traps to enhance photon collection. A WLS photon trap can be further enhanced by encapsulating the PMT and the WLS plate with a mirror box. This mirror box will utilize broadband mirrors to reflect light on the sides and the bottom of the WLS, and the box is completed by stretching a dichroic mirror above the PMT. By optimizing a dichroic mirror to reflect only the WLS's emission spectrum, Cherenkov radiation absorbed by the WLS will be reflected inside the photon trap after WLS emission. We trace the light path of each photon to investigate the individual and combined benefits of the dichroic mirror and WLS. We exclusively determine how each photon reached the PMT, using the classification scheme shown in figure~\ref{fig_setup} and summarized below:  

\begin{enumerate}
\item {\it Primary:} A photon hits the PMT directly.
\item {\it Internal:} A photon is absorbed in the WLS, re-emitted, and subsequently collected by the PMT while propagating only within the WLS.
\item {\it External:} A photon is absorbed in the WLS, re-emitted, and then collected by the PMT. The re-emitted photon would have escaped the trap without the presence of the dichroic mirror.
\end{enumerate} 

Relevant for the performance of a photon trap is its ability to detect photons. Therefore, we compute the detection efficiency~(DE), which is the product of the collection efficiency~(CE) and the corresponding quantum efficiency~(QE) of the PMT. The efficiencies are wavelength dependent and quoted efficiencies here are integrated over a Cherenkov spectrum from 200~nm to 700~nm. We quote collection efficiencies alongside with detection efficiencies to remove bias due to the choice of the PMT.

We demonstrate the functionality of a $100~{\rm cm} \times 100~{\rm cm}$ dichroic mirror box trap relying on a 12-inch PMT and show that it can provide competitive performance in total photon collection compared to a single 20-inch PMT. 
The improvements in collection efficiency come at the price of an inevitable increased time delay in secondary photon detection. 
We explore designs that minimize the time spread between direct Cherenkov detection (primary) and facilitated (secondary) photon detection due to the trap.\\

This report is structured as follows: In section~2 we introduce the photon trap designs, technical solutions and our simulation. In section~3 we discuss the results of our simulations and compare the performance of different trap configurations. Section~4 concludes and discusses possible applications and improvements of our design.\\


\section{Photon trap design and simulation} 

In this section we introduce the basic photon trap designs followed by a description of our simulation and physical properties of the materials as implemented in GEANT4~\cite{Agostinelli:2002hh}. 

\subsection{Photon trap configurations}

\begin{figure}[h]
\centering
\includegraphics[width=0.9\textwidth]{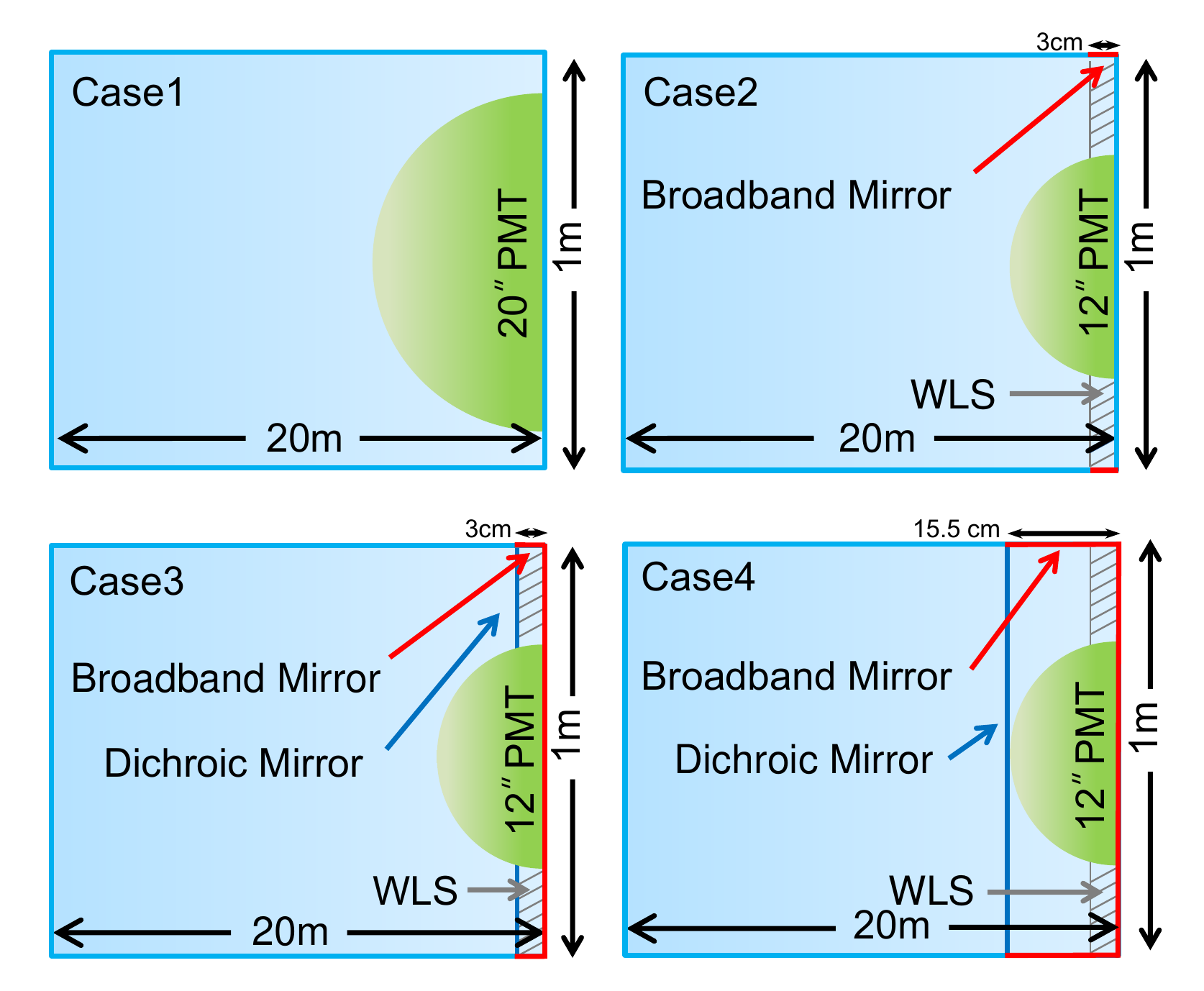}
\caption{Schematic drawings of the four different test configurations used for our study. PMTs are not to scale to emphasis the trap design. }
\label{fig_cases}
\end{figure}

First we introduce three photon trap designs, which distinguish themselves through their increasing complexity and compare to Hyper-K's baseline conventional design consisting of a single large (20-inch or 50.5~cm) PMT (case~1). Cases 2-4 use a 12-inch (30.5~cm) PMT combined with a WLS and enhanced photon trapping techniques. Unless otherwise specified the WLS plates have a thickness of 3~cm and are kept to one-by-one meter size with a centered 12-inch diameter hole to fit the PMT. A dichroic mirror is added on top of the WLS plate (case~3) and encases PMT and WLS (case~4). The photon trap designs are summarized in the following and schematic drawings are given in figure~\ref{fig_cases}.

\newpage

\begin{itemize}
\item {\bf Case 1: "Baseline":} The baseline design consists of a single 20-inch PMT. 
\item {\bf Case 2: "WLS trap":} A 12-inch PMT is surrounded by a WLS plate, that is covered with broadband mirrors on the sides. 
\item {\bf Case 3: "Mirror box":} A 12-inch PMT is surrounded by a WLS plate, that is covered with broadband mirrors on the bottom and sides. A dichroic mirror is on the top surface of the WLS. Bottom and top mirrors stretch one-by-one meter, and the dichroic mirror has a 12-inch-diameter hole at its center to fit the PMT.
\item {\bf Case 4: "Full trap": } A 12-inch PMT is surrounded by a WLS plate, that is covered with a broadband mirror on the bottom. A one-by-one meter dichroic mirror encases both the PMT together with the WLS and a side broadband mirror extends from the bottom to the dichroic mirror.
\end{itemize}

\subsection{Wavelength shifter}
\begin{figure}[h]
\centering
\includegraphics[width=0.8\textwidth]{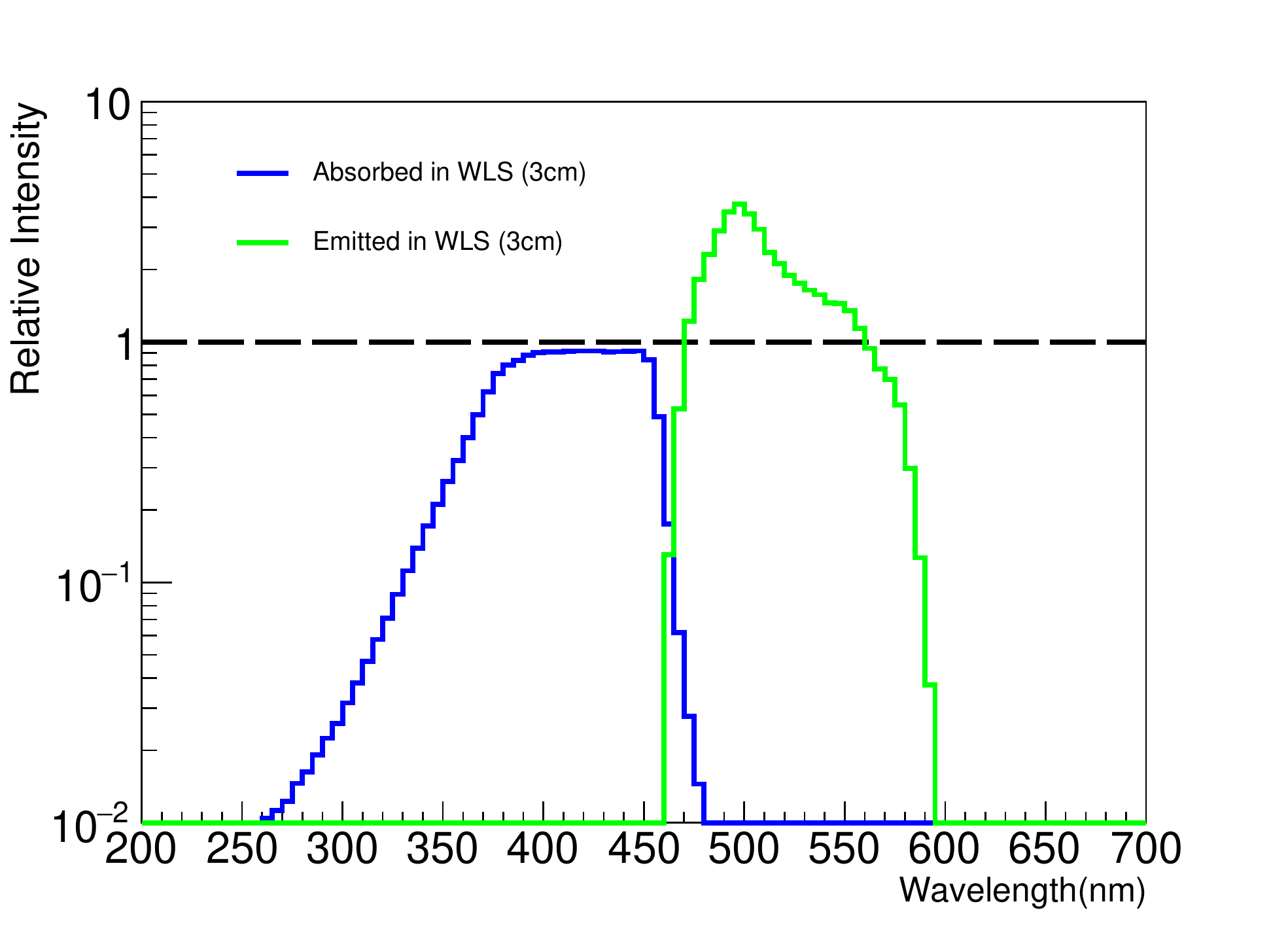}
\caption{Characteristics of the BC482-A wavelength shifter as implemented in GEANT4 with Cherenkov radiation injected orthogonal directly above a 3~cm thick WLS plate. The absorption and emission spectrum is normalized relative to the Cherenkov spectrum intensity at the corresponding wavelength.}
\label{fig_WLSspectra}
\end{figure}

For our study we use the widely available BC482-A blue-to-green WLS produced by Saint-Gobain Crystals~\cite{BC482A}. This WLS efficiently absorbs light between the 400-450~nm and dominantly emits between 475-525~nm. The absorption and emission spectra for a 3~cm thick WLS as implemented in GEANT4 simulation is shown in figure~\ref{fig_WLSspectra}. We generate Cherenkov radiation at the top of the WLS to minimize the effect of water absorption. GEANT4's optical process functionalities have provided us the ability to simulate optical properties of a WLS. We note that our simulation also allows for WLS-emitted photons to get absorbed and re-emitted.

The manufacturer specifies the average time delay between absorption and re-emission as 12~ns. The index of refraction of BC482-A ($n_{WLS}=1.59$) is higher compared to water ($n_{water}\approx1.35$) resulting in a critical angle of $58^{\circ}$, which leads to an effective trapping of isotropically re-emitted photons.

\subsection{Broadband mirror}

One of the elements used in our photon trap is a broadband mirror, which helps to capture photons that traverse the WLS without being absorbed. Side mirrors prevent re-emitted photons to escape to the sides thereby increasing the efficiency of the trap. They also have an important role in keeping the signal localized and not having photons detected in adjacent traps. We have implemented the broadband mirror in GEANT4 to have 98.5\% reflectivity for all relevant wavelengths (200~nm to 700~nm). The height of the broadband mirror is 3~cm, but we stretch it in case~4. 

\subsection{Dichroic mirror}

\begin{figure}[t]
\centering
\includegraphics[width=0.75\textwidth]{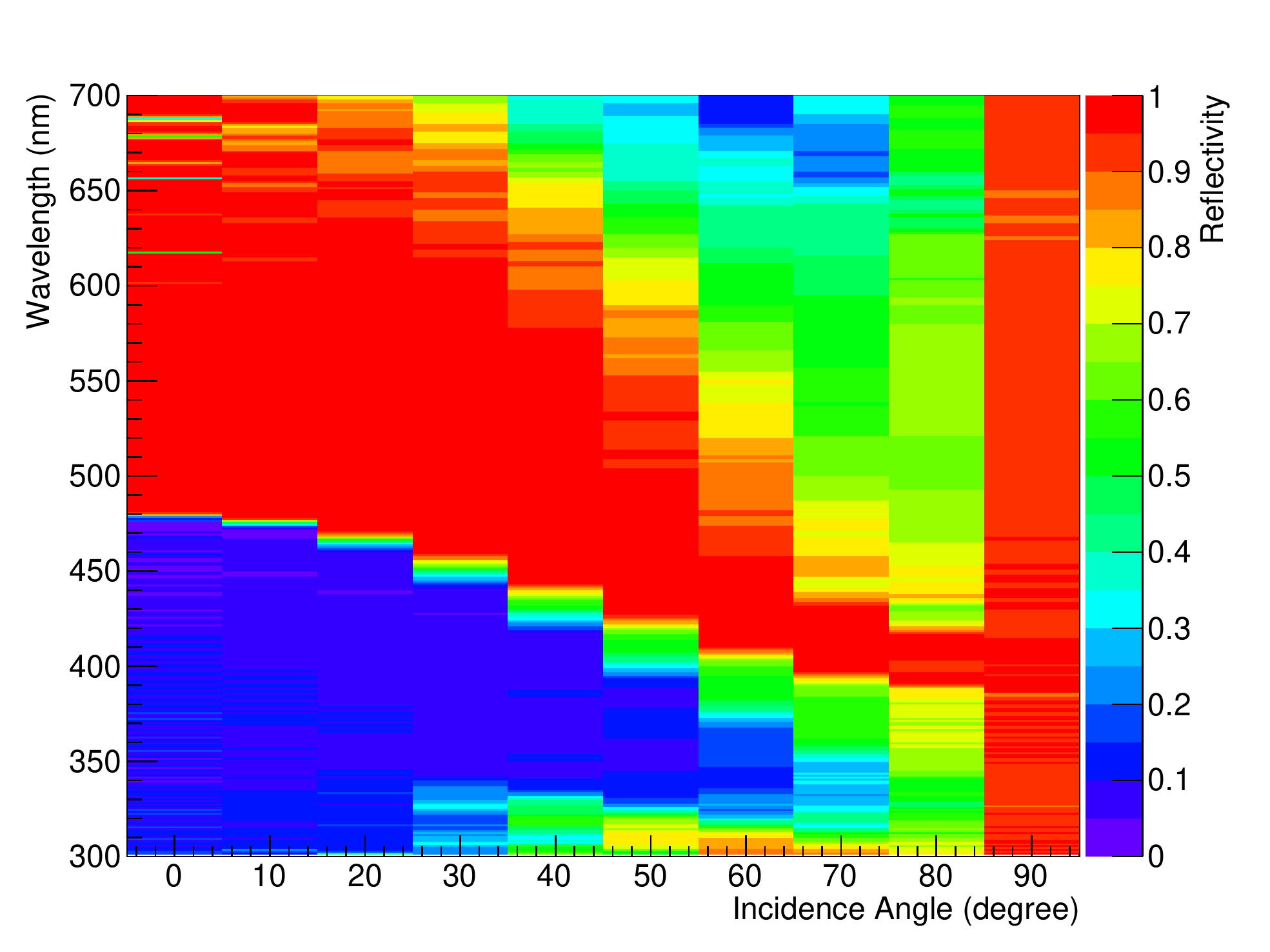}
\caption{Customized dichroic mirror reflectivity as provided by Iridian Spectral Technologies~\cite{Iridian} and used for our study.}
\label{fig_mirror}
\end{figure}

The dichroic mirror used in our simulations was customized for our purposes by Iridian Spectral Technologies~\cite{Iridian}. The key design feature of the dichroic mirror is that its reflectivity matches the emission spectrum of BC482-A, in particular at angles for which there is no total internal reflection within the WLS. As seen in figure~\ref{fig_mirror}, the mirror has high reflectivity in the WLS emission range (475-575~nm) at small incident angles. At angles greater than 50$^{\circ}\mathrm{}$ its reflectivity is somewhat reduced, yet it still remains at approximately 60\%. The mirror is transparent to incoming Cherenkov radiation at wavelength where the WLS is absorbing and where the PMT has high QE. 

\subsection{Photomultiplier tubes}

\begin{figure}[h]
\centering
\includegraphics[width=0.75\textwidth]{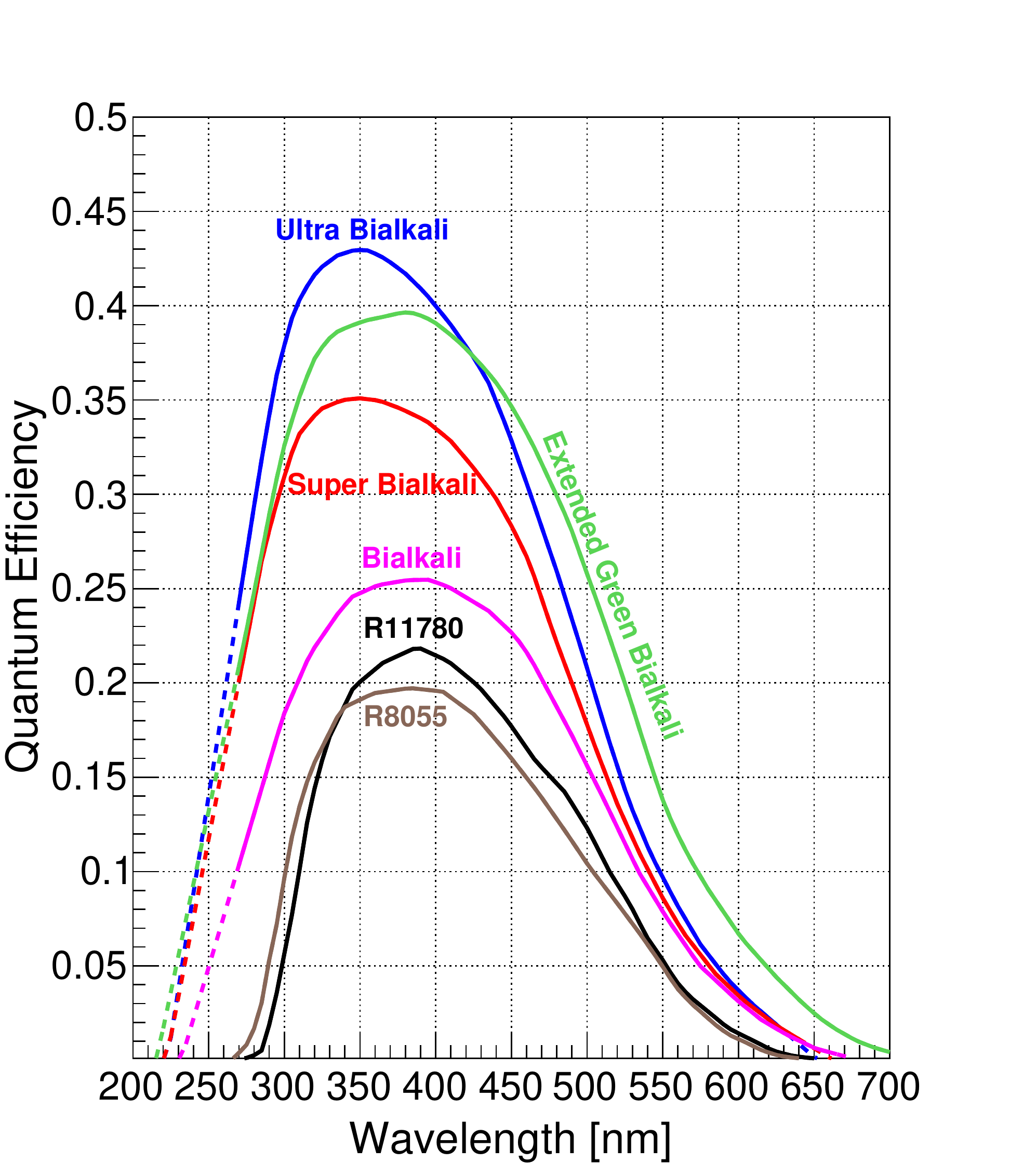}
\caption{Quantum efficiencies as used in our studies for Hamamatsu R11780, R8055 as well as Bialkali~(BA), Super Bialkali~(SBA), Ultra Bialkali~(UBA), Extended Green Bialkali~(EGB). Efficiencies were taken from manufacturer datasheets and extrapolated where needed~\cite{hamamatsu_datasheet}.}
\label{fig_PMT_QE}
\end{figure}

For our study we use a high quantum efficiency~(HQE) Hamamatsu Bialkali~(BA) PMT as reference PMT.  A PMT is needed to obtain the photon detection efficiency from the collection efficiency of a trap. Figure~\ref{fig_PMT_QE} shows our reference PMT QE compared to other candidates~\cite{hamamatsu_datasheet}. For easy comparison we applied QEs of different PMTs, without changing underlying PMT performance parameters. PMT performance was assumed to be similar to that of R11780, which has been extensively studied~\cite{Brack:2012ig} and found to be an excellent candidate PMT for very large-scale water Cherenkov or scintillator detectors. The single photo electron (SPE) charge and timing response were determined to be excellent, and the high quantum efficiency version increases the detected photon yield on average by about 50\% compared to the standard QE version. A two-dimensional scan of the HQE PMT revealed that the charge and timing response is uniform across most of the photocathode surface. Timing shifts near the edge of the PMT are less than 3~ns, due to an alternative dynode structure designed by Hamamatsu to mitigate large shifts observed in the standard configuration~\cite{Brack:2012ig}.  

\subsection{Simulation setup and  Cherenkov wavelength distribution}

\begin{figure}[h]
\centering
\includegraphics[width=0.9\textwidth]{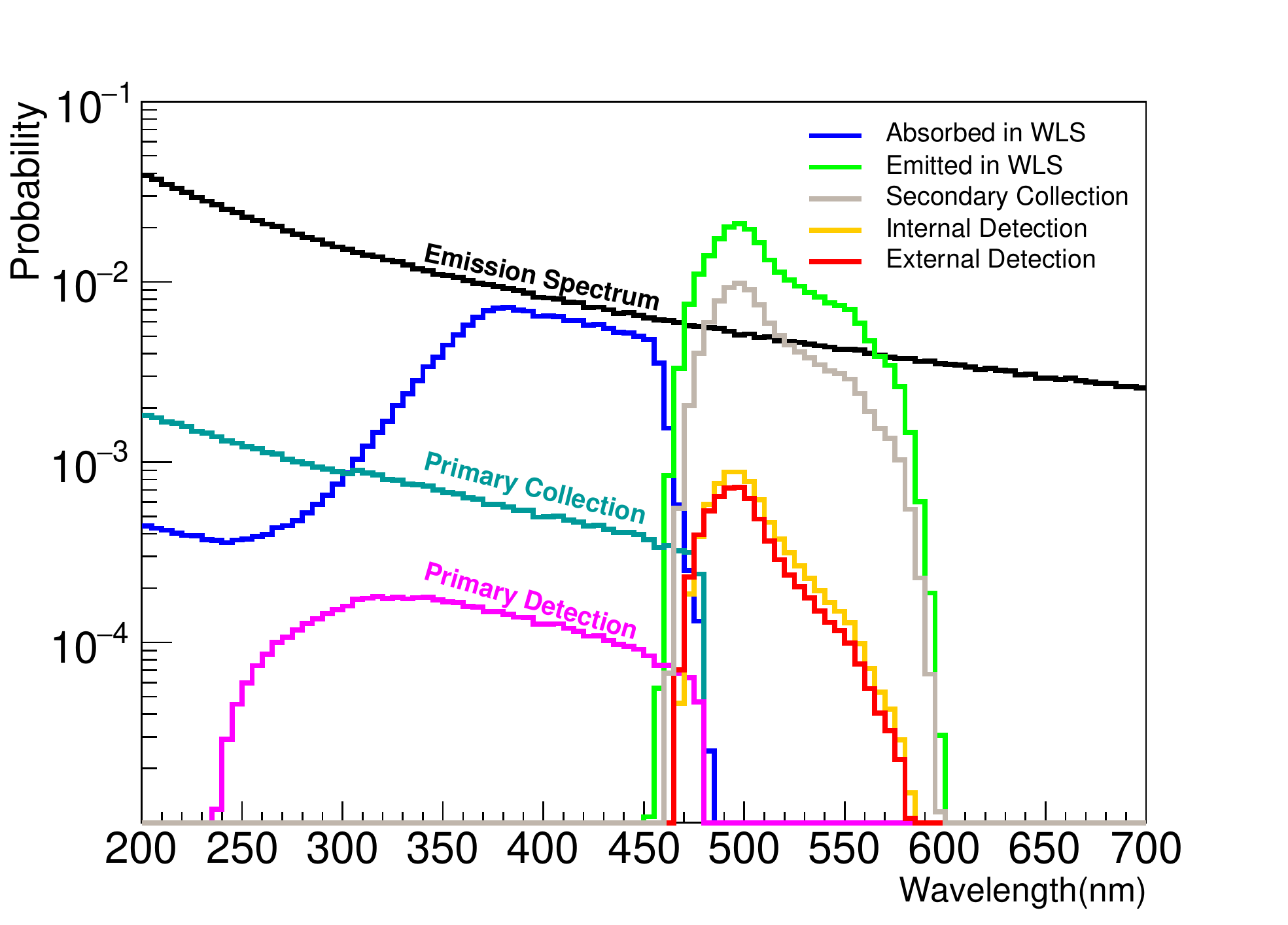}
\caption{Absorption, emission and detection spectra are shown following our classification scheme. The shown result is for the simulation of a full $100~{\rm cm} \times 100~{\rm cm}$ photon trap (case~4) with a 12-inch BA PMT and 3~cm thick WLS and dichroic mirror.}
\label{fig_TrapSpectra}
\end{figure}

In our simulation, we inject light sampled from the water Cherenkov spectrum between 200-700~nm (see figure~\ref{fig_TrapSpectra}) toward the photon trap under study. Radiation is generated 20~meters away from the photon trap, which is the approximate radius of a sphere fitted inside Hyper-K. The angle of the Cherenkov radiation is kept perpendicular to the trap's horizontal plane. The position of each Cherenkov photon is randomized within an one square meter area. For sufficient statistics we inject 10~million photons for each study.
Figure~\ref{fig_TrapSpectra} shows the injected Cherenkov spectrum and relative spectra for primary and secondary collected and detected photons. One can see that although the BC482-A WLS is optimized to absorb blue light, it also collects some amount of UV Cherenkov radiation. The amount of UV Cherenkov absorbed by the WLS increases with WLS thickness and a 15~cm thick WLS absorbs five times more than a 3~cm WLS. \\
Secondary emission spectra from the WLS are not identical for different trap configurations,
as our simulation allows for WLS-emitted photons to get absorbed and re-emitted. Due to energy conservation the wavelength of photons gets shifted to blue when this process is iterated. Figure~\ref{fig_TrapSpectra} shows WLS absorption and emission spectra as well as collection and detection spectra for a BA PMT.\\

\section{Results}
\label{sec_results}

In this section we report the performance results of each photon trap as evaluated with our GEANT4 simulation. We give collection efficiencies, which indicate how effectively photons can reach the PMT in the different designs. We compute detection efficiencies and timing using the wavelength dependent QE of a BA PMT and discuss the effect of using different PMTs. 

\subsection{Photon collection and detection efficiencies}

Collection efficiencies of the studied photon trap are compared in table~\ref{tab1}. Primary, internal, and external collection efficiencies are reported relative to the primary collection of a 20-inch PMT (case~1). The total efficiency is given by the sum of primary and secondary collection efficiencies. Photon detection efficiencies using a BA as benchmark PMT are given in table~\ref{tab5}. The statistical uncertainty on the collection (detection) efficiency is $0.1\%$ ($0.3\%$).
\begin{table}[h]
\centering
\caption{Comparison of the photon collection efficiencies for the tested configurations. Collection efficiencies are reported relative to our baseline case~1 (primary collection of a 20-inch PMT). The statistical uncertainty on the reported values is 0.1\%.\label{tab1}}
\begin{tabular}{|l|l|l|l|l|l|l|l|}
\hline
 & \multicolumn{4}{|c|}{\textbf{Relative collection efficiency}} \\ \hline
\textbf{Configuration} & \textbf{Primary} & \textbf{Internal} & \textbf{External} & \textbf{Total}\\ \hline
Case 1 & 1.000  	& 0 		& 0 		& 1.000 \\ \hline
Case 2 & 0.380  	& 0.321 	& 0		& 0.701 \\ \hline
Case 3 & 0.379 	& 0.355 	& 0.104  	& 0.838 \\ \hline
Case 4 & 0.306  	& 0.369  	& 0.302 	& 0.978 \\ \hline
\end{tabular}
\end{table}

\begin{table}[h]
\centering
\caption{Relative photon detection efficiency of tested configurations using the QE of a BA. The error on the reported efficiencies is 0.3\%.}
\label{tab5}
\begin{tabular}{|l|l|l|l|l|l|l|l|}
\hline
& \multicolumn{4}{|c|}{\textbf{Relative detection efficiency}} \\ \hline
\textbf{Configuration} & \textbf{Primary} & \textbf{Internal} & \textbf{External} & \textbf{Total}\\ \hline 
Case 1 & 1.000  	& 0 		& 0 		& 1.000 \\ \hline 
Case 2 & 0.379  	& 0.358  	& 0		& 0.737 \\ \hline
Case 3 & 0.378   	& 0.396	& 0.099 	& 0.874 \\ \hline
Case 4 & 0.316 	& 0.412 	& 0.344	& 1.071 \\  \hline
\end{tabular}
\end{table}

The positioning of the dichroic mirror between the Cherenkov source and the PMT leads to a reduction in primary collection efficiency, which is inevitable and apparent in case~4. However, with an increase in secondary collection, case~4 yields the highest detection efficiency.\\

\subsection{Trap size and positional dependence of time delay}

\begin{figure}[h]
\centering
\includegraphics[width=1.0\textwidth]{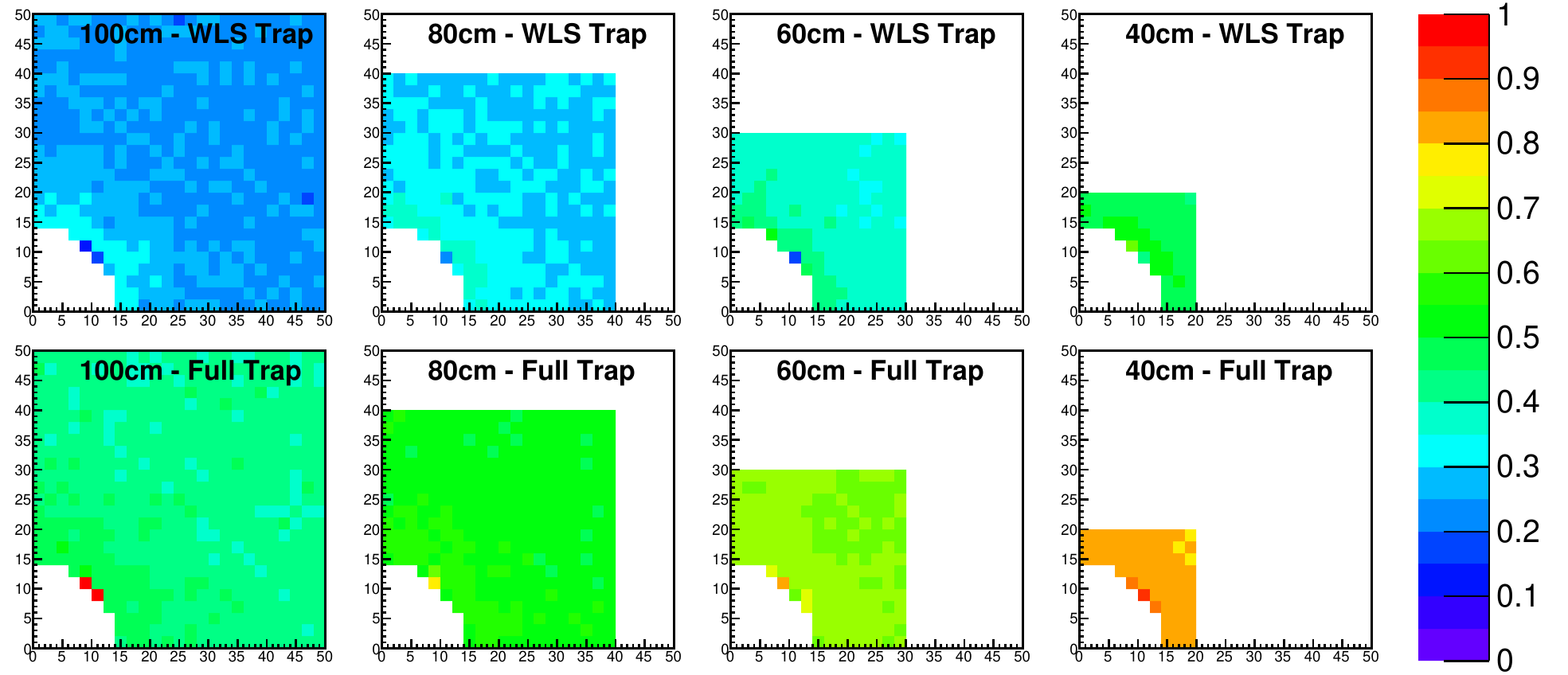}
\caption{Secondary photon (internal+external) collection efficiency maps of the WLS trap (case~2) and the full trap (case~4) relative to photons emitted from the WLS are shown in the top and bottom, row respectively. For symmetry reasons we only show a quarter of the trap with the bottom left side representing the PMT, where photons are collected.}
\label{fig_WLS_eff}
\end{figure}

We provide collection efficiency maps of a WLS trap (case~2) and full trap (case~4) for different square trap sizes ranging from $100 \times 100~{\rm cm}^2$ to $40 \times 40~{\rm cm}^2$ and explore the time delay of secondary to primary collection. 
The WLS thickness is kept at 3~cm and the PMT size is fixed to 12-inch. Figure~\ref{fig_WLS_eff} provides the secondary photon collection efficiencies depending on where the photon was initially emitted in the WLS. Each bin is relative to photons emitted from the WLS.\\

As expected, smaller traps have higher secondary photon collection efficiencies due to the shorter light path, however if the efficiency is normalized to a square meter unit area the overall collection efficiency is reduced. The shorter  geometrical light path results in reduced delay times as reported in table~\ref{tab3}. 
 
\begin{table}[h]
\centering
\caption{Comparison of the secondary photon collection and detection time delay between the WLS trap (case~2) and full trap (case~4). All times are given relative to the average collection (detection) time of the primary photons of the respective trap. The error on the mean and RMS is less than 0.03~ns.\label{tab3}}
\begin{tabular}{|c|c|c|c|c|}
\hline
 & \multicolumn{2}{|c}{\textbf{WLS Trap (case~2) Delay Time}}   & \multicolumn{2}{|c|}{\textbf{Full Trap (case~4) Delay Time}} \\ \hline
Size (${\rm cm}^2$)    & Mean (ns) & RMS (ns) & Mean (ns) & RMS (ns) \\ \hline
$100 \times 100$        & 22.07 (25.25)   & 20.56 (19.39)    & 25.37 (28.52)     & 22.11 (21.04)     \\ \hline
$80 \times 80$            & 17.90 (21.54)    & 18.41 (17.43)    & 20.76 (24.33)     & 19.82 (18.86)     \\ \hline
$60  \times 60$           & 13.54 (17.67)    & 15.91 (15.28)    & 15.62 (19.70)     & 16.94 (16.27)     \\ \hline
$40   \times 40$          & 9.08 (13.68)      & 13.30 (13.10)    &  10.32 (14.90)     & 13.74 (13.51)    \\ \hline
\end{tabular}
\end{table}

\begin{figure}[h]
\centering
\includegraphics[width=1.0\textwidth]{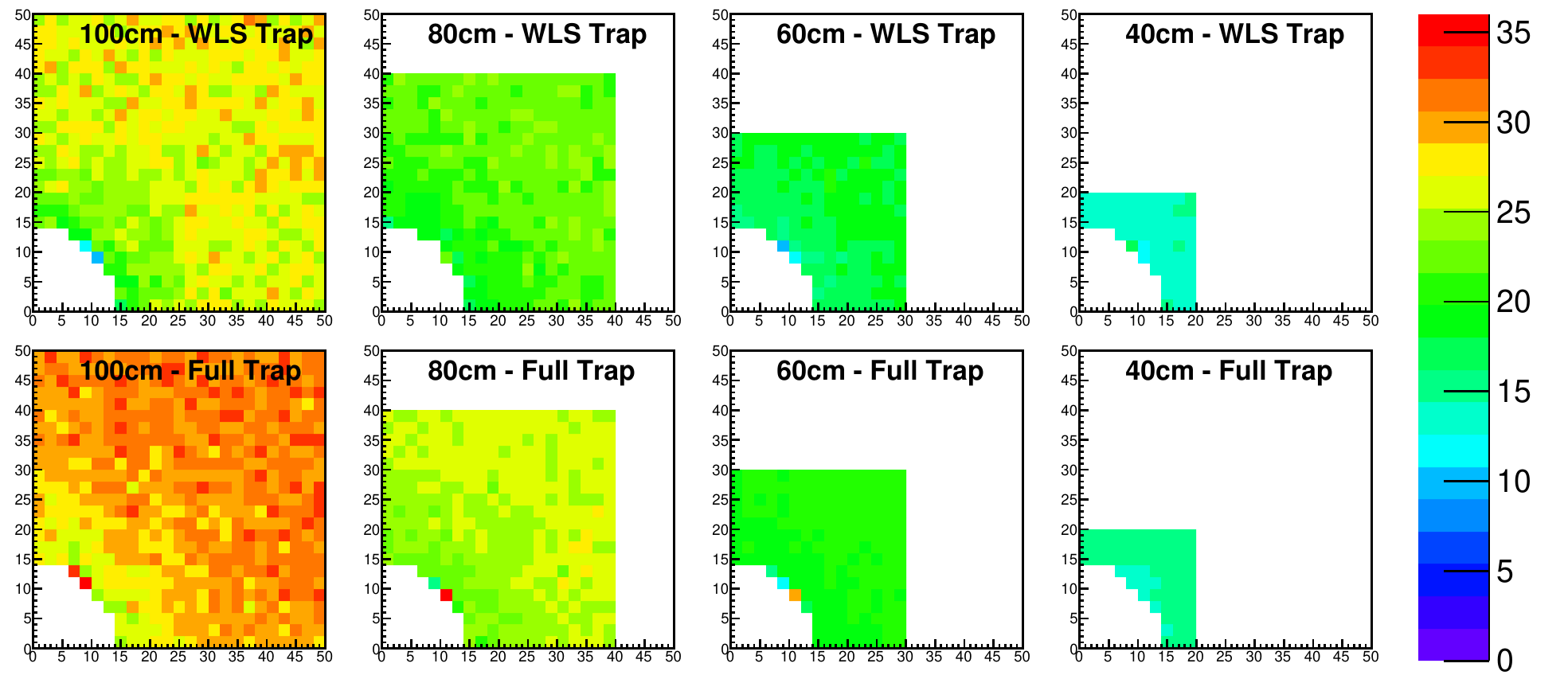}
\caption{Mean time delay in ns of secondary photons with respect to primary detected photons.}
\label{fig_WLS_mean_time}
\end{figure}

Figure~\ref{fig_WLS_mean_time} shows the average time delay of secondary photon detection relative to the mean primary detection time. The secondary photons originating further from the PMT show a higher mean time delay. The spread in mean delay time is minimal for the smallest trap. BC482-A  has a re-emission time of 12~ns~\cite{BC482A}, which is the dominant cause of delay for traps up to a size of about $80 \times 80~{\rm cm}^{2}$. We note that relative time delays of less than the 12~ns time constant are possible due to dispersion in the water and the different wavelength response of the PMT and the WLS.

\subsection{Timing resolution}

\begin{figure}[t]
\centering
\includegraphics[width=0.8\textwidth]{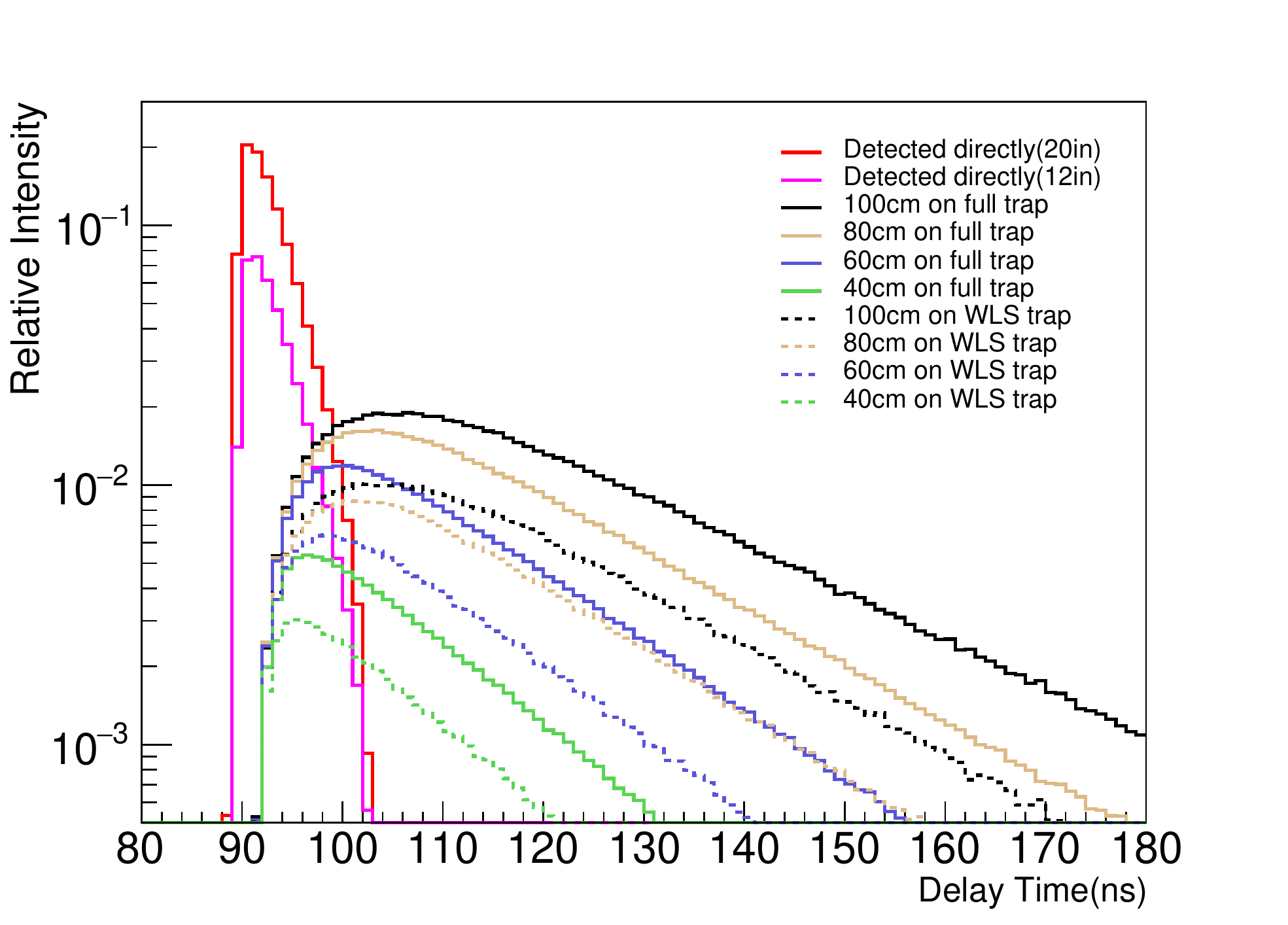}
\caption{Timing distribution of primary and secondary detected photons. The timing is reported relative to the injection time 20~m away from the photon trap and includes dispersion effects in the water that are of order ($\sim 3$~ns). "Detected directly" refers to the primary detection of a 20-inch PMT and all other distributions are normalized with respect to it.}
\label{fig_WLS_timing}
\end{figure}

\begin{figure}[h]
\centering
\includegraphics[width=0.49\textwidth]{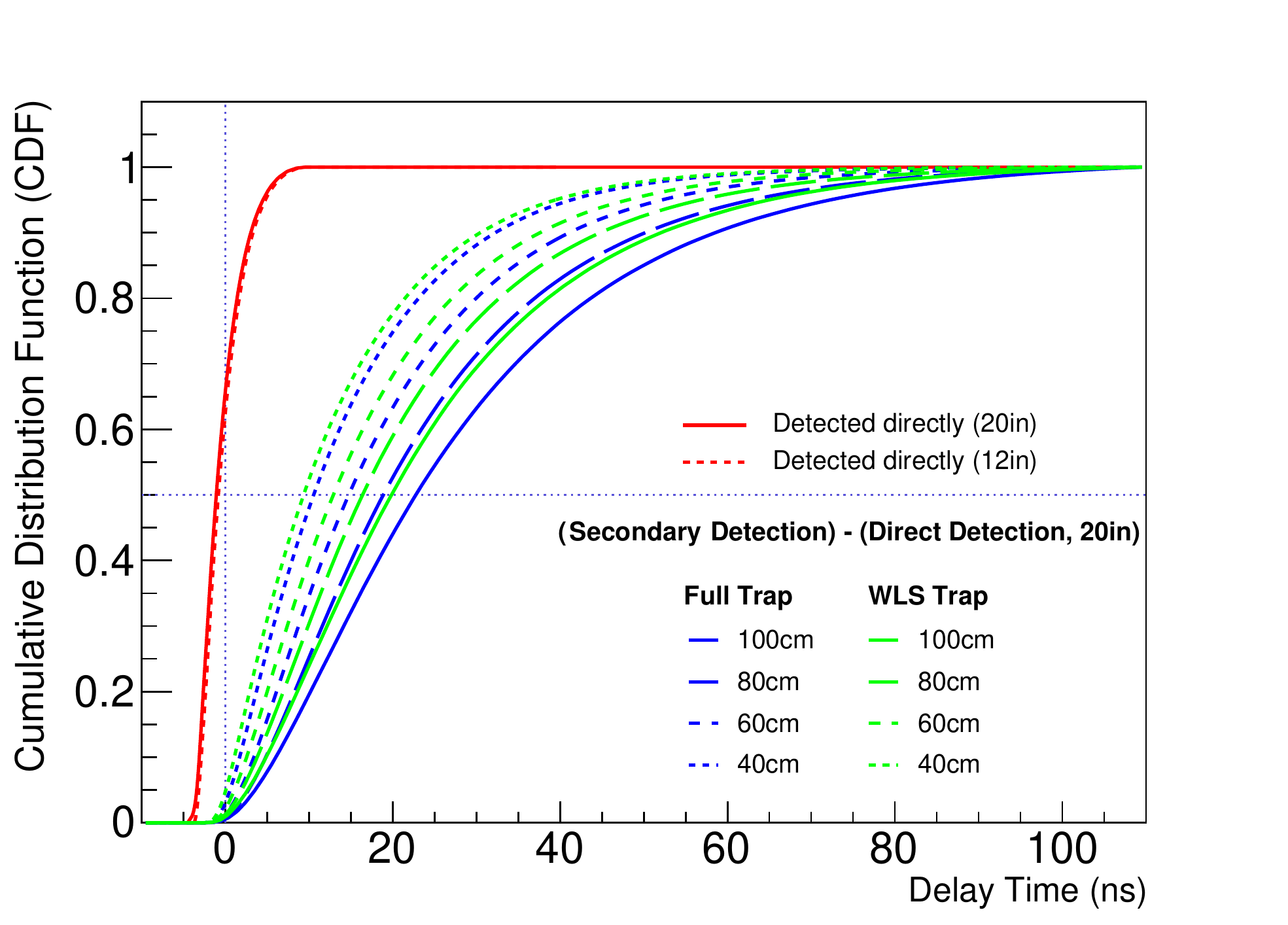}
\includegraphics[width=0.49\textwidth]{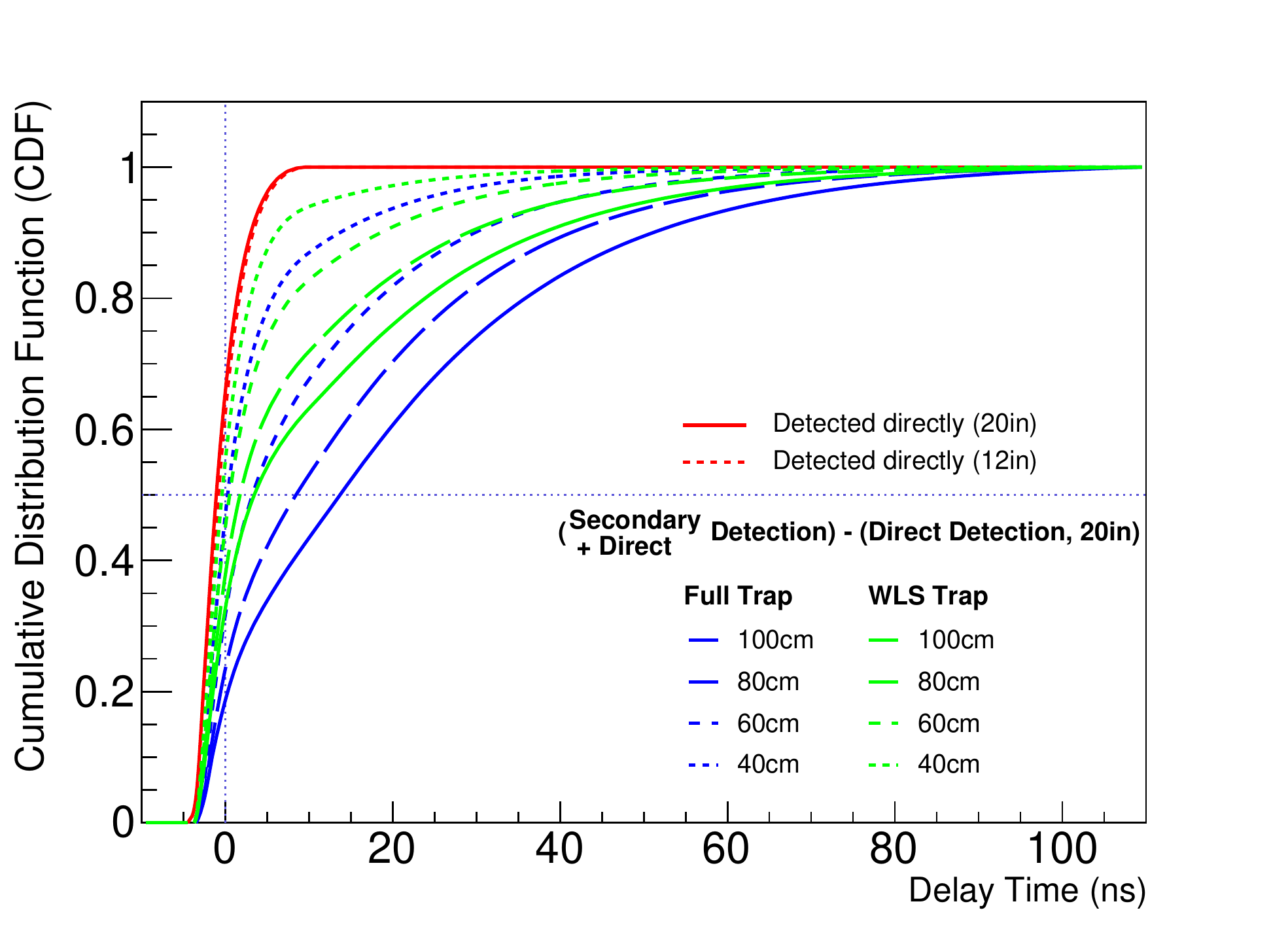}
\caption{Cumulative timing distributions of detected photons with the baseline design (20-inch PMT) compared to the WLS trap (case 2) (green) and the full trap (case 4) (blue). Left: Timing distribution exclusive by detection channel for both direct and secondary. Right: Timing distribution of the detected photons via all (direct+secondary) detection channels.}
\label{fig_cumulative}
\end{figure}

We compare the time delay of the secondary detection relative to the primary detection for different square trap sizes. Figure~\ref{fig_WLS_timing} provides a comparison of the primary and secondary detection of a WLS trap (case~2) and a full trap (case~4). The difference between delay of collected and detected photons originates from a higher fraction of UV photons collected directly by the PMT, below the PMTs sensitivity. To better understand the time delay encountered in the various configurations we study cumulative distributions of the detected photons within a certain time window (see figure~\ref{fig_cumulative}).

The RMS of the primary detection distribution in figure~\ref{fig_WLS_timing} is $2.42$~ns. As seen, both traps have a very wide spread of secondary photon detection time with RMS values of the order of 10~ns. In addition, the mean delay between primary and secondary photons is one order of magnitude larger than the time it takes a photon to travel one~meter, which will be the distance between two PMTs inside Hyper-K. While this clearly disfavours the usage of photon traps for the purpose of precision event reconstruction in Hyper-K's inner detector, advanced event reconstruction algorithms might still be able to provide the needed precision to achieve Hyper-K's physics goals. Algorithms developed at neutrino telescopes relying on unscattered light could offer interesting opportunities. The cumulative timing distribution shown in figure~\ref{fig_cumulative} shows that the majority of photons do not arrive with significant delays. We also note that the usage of the full trap introduces an additional mean time delay compared to the WLS trap of up to $\sim3$~ns. This additional delay can be considered to be a reasonable tradeoff for certain applications to increase the photon collection efficiency.
Although a significant delay in secondary photon detection relative to primary photons is clearly seen for the enhanced photon traps, the impact on the physics sensitivity due to the time delay is unclear and needs to be evaluated in a full detector simulation. 

\subsection{WLS thickness dependence}

After determining that case~4 had the best efficiency at standard WLS thickness (3~cm), we vary its thickness. As seen from table~\ref{tab2}, a thicker WLS clearly enhances the photon trap efficiency. In the case of a 15~cm thick WLS the efficiency is $20\%$ higher than a 20-inch PMT, however its timing resolution was worse with $\sim5$~ns. The increased efficiency of a thicker WLS can be mostly attributed to more absorption in the WLS with a consistent optical depth. Absorption in WLS is less efficient at the region of PMT sensitive wavelength, but it is enhanced in a thicker WLS.

\begin{table}[h]
\centering
\caption{Photon collection (detection) efficiency dependence on WLS thickness. Each efficiency is relative to case~1, respectively.  The statistical uncertainty on the reported collection and detection (in brackets) efficiencies are 0.1\% and 0.3\%, respectively.}
\label{tab2}
\begin{tabular}{|r|c|c|c|c|}
\hline
\textbf{Case 4} & \textbf{Primary} & \textbf{Internal} & \textbf{External} & \textbf{Total}\\ 
\textbf{Width of WLS} & \textbf{collection} & \textbf{collection} & \textbf{collection} & \textbf{collection} \\  \hline
5~mm     & 0.306 (0.315)	& 0.160 (0.166) & 0.354 (0.386)	& 0.820 (0.867) \\ 
30~mm   & 0.306 (0.316) 	& 0.369 (0.412) & 0.302 (0.344)	& 0.978 (1.071) \\
150~mm & 0.305 (0.315)  	& 0.488 (0.539)	 & 0.412 (0.389)	& 1.205 (1.242) \\ \hline
\end{tabular}
\end{table}

\subsection{PMT dependence}

We study the performance of different PMTs for our design and quote the ratios of total detection efficiency for various PMTs relative to a BA 20-inch setup (see table~\ref{tab6}). It can be seen that an Extended Green Bialkali (EGB) which matches our WLS well in the emission spectrum achieves the best detection efficiency, followed by Ultra Bialkali (UBA).  

\begin{table}[h]
\centering
\caption{Relative detection efficiency compared to case~1 with Bialkali (BA) as baseline.}
\label{tab6}
\begin{tabular}{|l|l|l|l|l|l|l|}
		\hline
		\textbf{Configuration} & \textbf{R8055} & \textbf{R11780} & \textbf{BA} & \textbf{SBA} & \textbf{EGB} & \textbf{UBA} \\ \hline
		Case~1 & 0.6097 & 0.6084 & 1 & 1.5311 & 1.7524 & 1.8422 \\ \hline
		Case~2 & 0.4705 & 0.5048 & 0.7371 & 0.9893 & 1.2575 & 1.1722 \\ \hline
		Case~3 & 0.5607 & 0.6074 & 0.8736 & 1.1437 & 1.4876 & 1.3503 \\ \hline
		Case~4 & \multicolumn{6}{|c|}{} \\ \hline
		5~mm WLS & 0.6519 & 0.6127 & 0.8671 & 1.1159 & 1.4687 & 1.3143 \\
		30~mm WLS & 0.6993 & 0.7717 & 1.0707 & 1.3500 & 1.8017 & 1.5869 \\
		150~mm WLS & 0.8085 & 0.8940 & 1.2418 & 1.5367 & 2.1009 & 1.7997 \\ \hline
\end{tabular}
\end{table}

\subsection{Angular dependence of the photon acceptance}

\begin{figure}[h]
	\centering
	\includegraphics[width=0.8\textwidth]{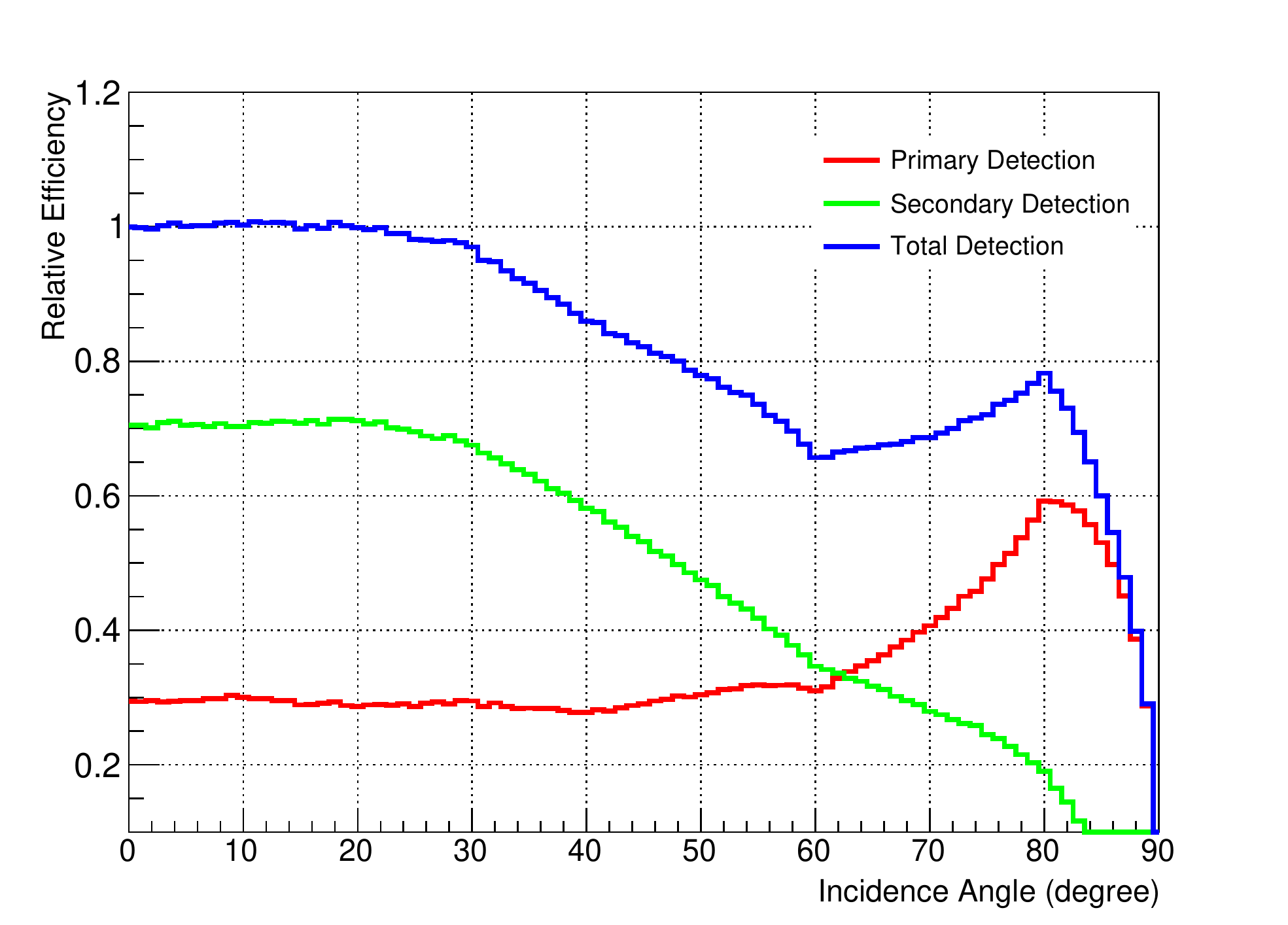}
	\caption{Relative detection efficiency of the full trap (case~4) for Cherenkov radiation injected just above the trap. The efficiency is normalized to one at $0^{\circ}$ incidence angle. Total efficiency is the sum of primary and secondary detection. The detailed behavior is described in the text.}
	\label{incident_dependency_efficient}
\end{figure}

We study the angular dependence of the photon trap acceptance by injecting photons directly above the trap at different angles. The angular dependence of the detection efficiency, normalized to one for orthogonal incidence angle, for the full trap (case 4) is reported in figure~\ref{incident_dependency_efficient}. 
Photons are injected just above the dichroic mirror to avoid any absorption of UV~light in the water. 
As can be seen in figure~\ref{incident_dependency_efficient} the secondary detection is flat up to $30^{\circ}$ and decreases for larger incident angles due to increased reflectivity of the dichroic mirror at relevant wavelengths. Primary and secondary detection are equal at $60^{\circ}$, after which primary detection gradually increases due to geometrical effects. At very large incident angles ($80^{\circ}$) light reflects of the dichroic mirror resulting in a steep decrease in efficiency. 

\section{Conclusion}

The photon collection efficiency of a large PMT can be matched by a smaller PMT in combination with a photon trap.
We have shown that a photon trap consisting of a 12-inch PMT, wavelength shifter, dichroic mirror, and broadband mirrors can achieve similar photon collection efficiencies as those of a 20-inch PMT at potentially lower cost. 
The detection efficiency of a photon trap can even surpass that of a single large PMT. Additionally, we note that a 12-inch PMT is expected to have a lower dark noise rate and better timing response compared to a 20-inch PMT. 
We have provided detailed comparisons of three photon trap designs relying on a 12-inch PMT and showed that the full trap (case 4) was able to detect more photons than a single 20-inch PMT. Detailed timing studies have been preformed and the delays due to re-emission and light propagation in the photon traps have been quantified.

In our study we injected light 20~m away from the photon trap, however absorption in the water significantly attenuates light with wavelength below 470~nm. UV light detectable by the WLS is expected to enhance trap efficiencies, making our result conservative. Further studies are needed that quantify the impact of UV light. A time delay due to dispersion in the water should also be noted. UV light which is most efficiently detected by the WLS takes about 3~ns longer to reach the photon trap compared to wavelengths closer to the PMT peak sensitivity, introducing a time delay, which is reduced if light is injected at shorter distances.

Our work has shown that enhanced PMT photon traps can achieve collection efficiencies competitive to single large PMTs. Traps designs could potentially be further improved with dedicated studies. For example, using a WLS (or multiple WLSs) which will shift both UV and blue light to green, could increase the collection efficiencies. Furthermore, utilizing faster WLSs will improve the timing resolution of photon traps. Using a different WLS might require changing the dichroic mirror to have a functioning photon trap. In addition, an improved trap geometry using for example a Winston cone could improve collection efficiencies further and achieve better timing resolution as re-emitted photons can be guided to the PMTs on a shorter path. Trap size studies indicate an improved timing resolution with smaller traps at the cost of reduced total photon collection. We focused on light injected perpendicular to the trap and observed that at an incident angle of $30^{\circ}$ and $60^{\circ}$ the detection efficiency is decreased by a factor of 0.99 and 0.67 respectively compared to normal incidence for the full trap.

Disadvantages of the photon trap are associated with a degraded timing resolution and reflected light on the trap.
Reflected primary photons contribute to the overall PMT count rates in a detector and might reduce sensitivity to very low energy events. The fraction of the reflected Cherenkov radiation is dependent on the incident angle and wavelength (for perpendicular incoming Cherenkov radiation, 28\% of the photons are reflected).  

While timing requirements depend on the physics objective, the most likely application of our photon trap would be in the outer (veto) detector array of large water Cherenkov detectors with less stringent timing requirements. Further studies on the physics impact of the reflected light is needed and if advanced reconstruction algorithms could mediate the impact. We note that wavelengthshifting optical sensor modules have also been proposed for DUNE~\cite{Whittington:2015rkr} and next generation neutrino telescopes~\cite{Hebecker:2015jyb,Aartsen:2014oha,Pfeiffer_ICRC2017}.

We note here that our simulation and design has focused on the photon trap efficiency and performance by itself. Engineering and safety modification such as encasing PMTs in fiber-reinforced plastic (FRP) with acrylic front windows~\cite{Abe:2013gga} to make designs shock wave resistant, need to be investigated for a full detector design. Further the enhanced PMT photon traps require full cost analysis to weigh their benefits.\\

\acknowledgments

We would like to thank the Hyper-Kamiokande proto collaboration for suggestions and discussions. We especially like to thank Yasuhiro Nishimura, F{\i}rat Nur\"{o}zler, Peter Peiffer and Hiro A. Tanaka for their comments and suggestions on this work. We would like to thank Iridian Spectral Technologies (Ottawa, Canada) for providing a realistic calculation of the dichroic mirror reflectivity.
F.~Retiere and P.~Gumplinger acknowledges support from TRIUMF and from the National Science and Engineering Research Council (NSERC) of Canada. C.~Rott acknowledges support from the Korea Neutrino Research Center which is established by the National Research Foundation of Korea~(NRF) grant funded by the Korea government~(MSIP) (No. 2009-0083526) and Basic Science Research Program (NRF-2016R1D1A1B03931688). S.~In is supported by Global PH.D Fellowship Program through the National Research Foundation of Korea (NRF) funded by the Ministry of Education (NRF-2015H1A2A1032363).


\begin{thebibliography}{99}

\bibitem{Fukuda:2002uc} 
  Y.~Fukuda {\it et al.} [Super-Kamiokande Collaboration],
  ``The Super-Kamiokande detector,''
  Nucl.\ Instrum.\ Meth.\ A {\bf 501}, 418 (2003).

\bibitem{HK_Design} 
  K.~Abe {\it et al.} [Hyper-Kamiokande Proto- Collaboration],
  ``Hyper-Kamiokande Design Report,''
 KEK Preprint 2016-21, ICRR-Report-701-2016-1, available at \href{https://lib-extopc.kek.jp/preprints/PDF/2016/1627/1627021.pdf}{https://lib-extopc.kek.jp/preprints/PDF/2016/1627/1627021.pdf}
  
\bibitem{Abe:2015zbg} 
  K.~Abe {\it et al.} [Hyper-Kamiokande Proto- Collaboration],
  ``Physics potential of a long-baseline neutrino oscillation experiment using a J-PARC neutrino beam and Hyper-Kamiokande,''
  PTEP {\bf 2015}, no. 5, 053C02 (2015)
  [arXiv:1502.05199 [hep-ex]].

  
\bibitem{Abe:2011ts} 
  K.~Abe {\it et al.},
  ``Letter of Intent: The Hyper-Kamiokande Experiment --- Detector Design and Physics Potential ---,''
  arXiv:1109.3262 [hep-ex].

\bibitem{T2HKK} 
  K.~Abe  {\it et al.} [Hyper-Kamiokande proto- Collaboration],
  ``Physics Potentials with the Second Hyper-Kamiokande Detector in Korea,''
  arXiv:1611.06118 [hep-ex].


\bibitem{Beacom:2010kk} 
  J.~F.~Beacom,
  ``The Diffuse Supernova Neutrino Background,''
  Ann.\ Rev.\ Nucl.\ Part.\ Sci.\  {\bf 60}, 439 (2010)
  [arXiv:1004.3311 [astro-ph.HE]].


\bibitem{Beacom:2003nk} 
  J.~F.~Beacom and M.~R.~Vagins,
  ``GADZOOKS! Anti-neutrino spectroscopy with large water Cherenkov detectors,''
  Phys.\ Rev.\ Lett.\  {\bf 93}, 171101 (2004)
  [hep-ph/0309300].


\bibitem{Hirata:1987hu} 
  K.~Hirata {\it et al.} [Kamiokande-II Collaboration],
  ``Observation of a Neutrino Burst from the Supernova SN 1987a,''
  Phys.\ Rev.\ Lett.\  {\bf 58}, 1490 (1987).


\bibitem{Ikeda:2007sa} 
  M.~Ikeda {\it et al.} [Super-Kamiokande Collaboration],
  ``Search for Supernova Neutrino Bursts at Super-Kamiokande,''
  Astrophys.\ J.\  {\bf 669}, 519 (2007)
  [arXiv:0706.2283 [astro-ph]].

\bibitem{Rott:2012qb} 
  C.~Rott, J.~Siegal-Gaskins and J.~F.~Beacom,
  ``New Sensitivity to Solar WIMP Annihilation using Low-Energy Neutrinos,''
  Phys.\ Rev.\ D {\bf 88}, 055005 (2013)
  [arXiv:1208.0827 [astro-ph.HE]].

\bibitem{Bernal:2012qh}
  N.~Bernal, J.~Martin-Albo and S.~Palomares-Ruiz,
  ``A novel way of constraining WIMPs annihilations in the Sun: MeV neutrinos,''
  JCAP {\bf 1308}, 011 (2013)
  [arXiv:1208.0834 [hep-ph]].

\bibitem{Rott:2015nma}
  C.~Rott, S.~In, J.~Kumar and D.~Yaylali,
  ``Dark Matter Searches for Monoenergetic Neutrinos Arising from Stopped Meson Decay in the Sun,''
  JCAP {\bf 1511}, no. 11, 039 (2015)
  [arXiv:1510.00170 [hep-ph]].

\bibitem{Murase:2016nwx} 
  K.~Murase and I.~M.~Shoemaker,
  Phys.\ Rev.\ D {\bf 94}, no. 6, 063512 (2016)
  [arXiv:1606.03087 [hep-ph]].

\bibitem{Shiozawa:1998si} 
  M.~Shiozawa {\it et al.} [Super-Kamiokande Collaboration],
  ``Search for proton decay via p ---> e+ pi0 in a large water Cherenkov detector,''
  Phys.\ Rev.\ Lett.\  {\bf 81}, 3319 (1998)
  [hep-ex/9806014].

\bibitem{Fukuda:1998mi} 
  Y.~Fukuda {\it et al.} [Super-Kamiokande Collaboration],
  ``Evidence for oscillation of atmospheric neutrinos,''
  Phys.\ Rev.\ Lett.\  {\bf 81}, 1562 (1998)
  [hep-ex/9807003].


\bibitem{Rott:2015kwa} 
  C.~Rott, A.~Taketa and D.~Bose,
  Scientific Reports 5, Article number: 15225 (2015)
  [arXiv:1502.04930 [physics.geo-ph]].


\bibitem{Hosaka:2005um} 
  J.~Hosaka {\it et al.} [Super-Kamiokande Collaboration],
  ``Solar neutrino measurements in super-Kamiokande-I,''
  Phys.\ Rev.\ D {\bf 73}, 112001 (2006)
  [hep-ex/0508053].

\bibitem{Cravens:2008aa} 
  J.~P.~Cravens {\it et al.} [Super-Kamiokande Collaboration],
  ``Solar neutrino measurements in Super-Kamiokande-II,''
  Phys.\ Rev.\ D {\bf 78}, 032002 (2008)
  [arXiv:0803.4312 [hep-ex]].
  
\bibitem{Abe:2010hy} 
  K.~Abe {\it et al.} [Super-Kamiokande Collaboration],
  ``Solar neutrino results in Super-Kamiokande-III,''
  Phys.\ Rev.\ D {\bf 83}, 052010 (2011)
  [arXiv:1010.0118 [hep-ex]].

\bibitem{Botner:1980uf} 
  O.~Botner {\it et al.},
  ``A Hadron Calorimeter With Wavelength Shifter Readout,''
  Nucl.\ Instrum.\ Meth.\  {\bf 179}, 45 (1981).

\bibitem{Balka:1987ty} 
  L.~Balka {\it et al.} [CDF Collaboration],
  ``The CDF Central Electromagnetic Calorimeter,''
  Nucl.\ Instrum.\ Meth.\ A {\bf 267}, 272 (1988).

\bibitem{Agostinelli:2002hh} 
  S.~Agostinelli {\it et al.} [GEANT4 Collaboration],
  ``GEANT4: A Simulation toolkit,''
  Nucl.\ Instrum.\ Meth.\ A {\bf 506}, 250 (2003).


\bibitem{BC482A} 
Saint-Gobain Crystals. BC482-A Wavelength Shifting Bars, . URL
http://www.crystals.saint-gobain.com/uploadedFiles/SG-Crystals/Documents/SGC\%20BC482A\%20and\%20BC484\%20Data\%20Sheet.pdf

\bibitem{Iridian}
Mirror properties provided by Iridian Spectral Technology (Ottawa, Ontario, Canada). http://www.iridian-optical-filters.com/


\bibitem{Brack:2012ig} 
  J.~Brack {\it et al.},
  ``Characterization of the Hamamatsu R11780 12 inch Photomultiplier Tube,''
  Nucl.\ Instrum.\ Meth.\ A {\bf 712}, 162 (2013)
  [arXiv:1210.2765 [physics.ins-det]].

\bibitem{hamamatsu_datasheet}
Hamamatsu Datasheet URL
http://www.hamamatsu.com/resources/pdf/etd/LARGE\_AREA\_PMT\_TPMH1286E.pdf
https://www.hamamatsu.com/resources/pdf/etd/p-dev\_2015\_TOTH0023E.pdf

\bibitem{Abe:2013gga} 
  K.~Abe {\it et al.},
  ``Calibration of the Super-Kamiokande Detector,''
  Nucl.\ Instrum.\ Meth.\ A {\bf 737}, 253 (2014)
  [arXiv:1307.0162 [physics.ins-det]].


\bibitem{Whittington:2015rkr} 
  D.~Whittington,
  JINST {\bf 11}, no. 05, C05019 (2016)
  [arXiv:1511.06345 [physics.ins-det]].

\bibitem{Hebecker:2015jyb} 
  D.~Hebecker {\it et al.},
  ``Progress on the development of a wavelength-shifting optical module,''
  PoS ICRC {\bf 2015}, 1134 (2016).

\bibitem{Pfeiffer_ICRC2017}
P.~Pfeiffer 
  ``Overview and Performance of the Wavelength-shifting Optical Module (WOM) for IceCube-Gen2,''
  PoS(ICRC2017)1052.

\bibitem{Aartsen:2014oha} 
  M.~G.~Aartsen {\it et al.} [IceCube PINGU Collaboration],
  ``Letter of Intent: The Precision IceCube Next Generation Upgrade (PINGU),''
  arXiv:1401.2046 [physics.ins-det].

\end{thebibliography}
\end{document}